\documentstyle[preprint,aps,epsf]{revtex}
\tightenlines
\def\Journal#1#2#3#4{{#1} {\bf #2}, #3 (#4)}

\def\IJMPB{{\em Int. J. Mod. Phys.} B}

\def\NPB{{\em Nucl. Phys.} B}

\def\PLB{{\em Phys. Lett.} B}

\def\PRA{{\em Phys. Rev.} A}

\def\PRD{{\em Phys. Rev.} D}
\def\PREP{{\em Phys. Rep.}}

\newcommand{\be}{\begin{equation}}
\newcommand{\ee}{\end{equation}\noindent}
\newcommand{\eei}{\end{equation}}
\newcommand{\bea}{\begin{eqnarray}}
\newcommand{\eea}{\end{eqnarray}\noindent}
\newcommand{\eeai}{\end{eqnarray}}

\newcommand{\hf} {{1\over2}}

\def\la{\lambda}
\def\eq#1{(\ref{#1})}

\def\ord{{\cal O}}
\def\ra{\rangle}
\def\la{\langle}

\begin{document}
\title{Tree-level renormalization}

\author{Jean Alexandre$^a$\thanks{alexandr@lpt1.u-strasbg.fr},
Janos Polonyi$^{a,b}$\thanks{polonyi@fresnel.u-strasbg.fr}}
\address{$^a$Laboratory of Theoretical Physics, Louis Pasteur University\\
3 rue de l'Universit\'e 67087 Strasbourg, Cedex, France}
\address{$^b$Department of Atomic Physics, L. E\"otv\"os University\\
P\'azm\'any P. S\'et\'any 1/A 1117 Budapest, Hungary}
\date{\today}
\maketitle
\begin{abstract}
It is shown in the framework of the $N$-component scalar model
that the saddle point structure may generate non-trivial 
renormalization group flow. The spinodal phase separation
can be described in this manner and a flat action
is found as an exact result which is valid up to any order of the
loop expansion. The correlation function
is computed in a mean-field approximation.
\end{abstract}

\section{Introduction}
The strategy of the renormalization group method is a successive
elimination of the degrees of freedom, a gradual
simplification of the dynamical systems \cite{wilson}. 
In Statistical
and Quantum Physics the quantity to consider is the partition
function or the path integral, and the method consists of 
the successive integration over the dynamical variables.
The renormalization group is usually viewed in this manner as a 
technical device to sum up the contributions of the
thermal or the quantum fluctuations. The result, the
renormalized trajectory is used to assess the importance 
and the effects of the fluctuations of a given scale.
The goal of this paper is to show that the renormalization 
group can be non-trivial even in the absence of the 
fluctuations. The fluctuations and the mean values 
are the easiest to separate by means of the saddle-point
expansion. It will be shown that the saddle 
point can induce a non-trivial renormalization,
independently of the fluctuations.

In order to make our point clearer consider first a lattice model
with topological defects,
\be\label{partfc}
Z=\prod\limits_x\int d\phi(x)e^{-{1\over\hbar}S[\phi(x)]}.
\ee
The topological defects of the size $\ell a$ 
($a$ is the lattice spacing) minimize the 
action and the tree level $\ord(\hbar^0)$ contributions to the
saddle point approximation can be written as a grand 
canonical partition function of the topological defects 
\cite{mini}. Suppose that we perform in dimension $d$ a real-space 
blocking, $a\to na$ with $n>\ell$,
\be\label{rspbl}
e^{-{1\over\hbar}\tilde S[\tilde\phi(\tilde x)]}=
\prod\limits_x\int d\phi(x)
\prod\limits_{\tilde x}\delta\left(\tilde\phi(\tilde x)-
{1\over n^d}\sum\limits_{x\in\tilde x}\phi(x)\right)
e^{-{1\over\hbar}S[\phi(x)]}.
\ee
The fraction of the topological defects which are on the
surface of the blocks is supressed by $\ell/n$. Most of
the topological defects are inside of a block and they can be
thought as the saddle point configurations in the elimination
of the variables of the given block. As long as the
topological defects are present the blocking has a tree-level,
non-fluctuating contribution. Naturally the inhomogeneous
saddle points break external symmetries which provide
plenty of zero modes. The integration over them
captures some effects of the fluctuations. The difference 
between the tree-level contribution and the genuine fluctuations
might seem to be reduced to a technical issue at this point. 
But the coupling
constant dependence is always more complicated and singular 
at the tree-level than it is in the loop contributions.
Thus the scaling laws extracted from the tree- or the loop-level
terms might be radically different \cite{mini,af}. 

We discuss in this work the tree-level renormalization effects of an 
$N$-component scalar field  model defined by the action 
$S_\Lambda[\vec\phi]$. It is useful to introduce
a constrained partition function given by
\be\label{cmod}
Z_\Phi=\int{\cal D}[\vec\phi]\delta\left(
{1\over V_d}\left|\int d^dx\vec\phi(x)\right|-\Phi\right)
e^{-{1\over\hbar}S_\Lambda[\vec\phi]},
\ee
where $V_d$ is the space-time volume. The effective potential
of the model \eq{partfc} is
\be
V_{eff}(\Phi)=-{\hbar\over V_d}\ln Z_\Phi.
\ee
The regulator of the path integral, the cutoff $\Lambda$ in the
momentum space, is not 
written explicitely. We select the parameters
of the action in such a manner that the $O(N)$ symmetry is spontaneously
broken in the vacuum. As $\Phi$ is chosen to be smaller
than the amplitude of the field in the vacuum,
$\Phi<\la|\vec\phi(x)|\ra$, one expects metastable or
unstable behavior for certain modes. We shall find
non-trivial saddle points in this case during the
blocking. The saddle point expansion actually
offers a systematical treatment of the metastable 
phase or the spinodal instabilities. The resulting effective action
for the unstable phase is found to be flat. It is remarkable 
that this result goes beyond the tree-approximation and 
remains valid to every order of $\hbar$.
Domain walls are reproduced as the saddle points in the spinodal 
phase separation and we obtain the Maxwell-construction for the
free energy, $V_{eff}(\Phi)$. A similar problem has already
been discussed for $N=1$ by keeping track of simple saddle
points numerically in ref. \cite{iir}. The present paper
generalises the results presented there and relies on
analytical methods, based on continuity assumptions
infered from the numerical study.

The saddle point approximation is used in two different manners
in the paper. This is because we use the differential form of the
renormalization group equations and we have two small parameters
at our disposal.
One is $\hbar$ to organize the loop-expansion and the other is
the fraction of the modes to be eliminated in a blocking step,
$\epsilon=(1-\ell/n)^d$ in the example mentioned above.
The usual saddle point approximation is recovered when the limit
$\hbar\to0$ is taken at finite $\epsilon$. This is sufficient
to provide a well defined renormalization group flow. But the minimum
of the action becomes highly degenerate
when the infinitesimal form of the
renormalization group method is employed, i.e. the limit $\epsilon\to0$ 
is made before $\hbar\to0$. The disadvantage of the 
degeneracy of the effective action is that the saddle point approximation 
is in general spoiled. This is because the minimum at the saddle point
is shallow, the curvature of the action being $\ord(\epsilon)$.
But the determination of the effective action remains reliable 
since $\epsilon$ not only makes the action flat but enters as a 
supression factor for the loop corrections, as well. Such a double
role of $\epsilon$ is behind an unusual feature of the saddle
point expansion: the appearence of fluctuations in the order 
$\ord(\hbar^0)$.
The advantage of the degeneracy is an enormous simplification
of the functional integration which is demonstrated
by the mean-field computation of the correlation function. The fluctuations 
arising in $\ord(\hbar^0)$ have no restoring force to the equilibrium
position, a characteristic feature of the mixed phase of the first
order phase transitions.

The organization of the paper is the following. Section II.
introduces the infinitesimal renormalization step 
which provides a more powerful version
of the renormalization group equation. A simple and important
property of the renormalized action in the unstable
region is presented in
Section III for a scalar model. The renormalized trajectory
is given in Section IV and Section V contains a brief
analysis of the correlation function obtained on the
tree-level. Finally, Section VI is for the summary.

\section{Infinitesimal renormalization}
We introduce the infinitesimal renormalization group
in this Section. The traditional Kadanoff-Wilson blocking 
transformations in the Fourier space is the lowering of the
cut-off $k\to k-\delta k$,
\be\label{kw}
e^{-\frac{1}{\hbar}S_{k-\delta k}[\vec\phi]}=
\int{\cal D}[\vec\psi]e^{-\frac{1}{\hbar}S_k[\vec\phi+\vec\psi]},
\ee
where $\vec\phi$ has Fourier components for $|p|\le k-\delta k$ and
$\vec\psi$ for $k-\delta k< |p|\le k$ \cite{wilson}. 
This blocking is infinitesimal if the number of modes 
eliminated is small compared to the number of the left over 
modes in the system,
\be
\epsilon={\delta k\over k}<<1.
\ee
The one-loop approximation to the blocking \eq{kw} yields the 
exact renormalization group equations as
$\epsilon\to0$ \cite{wh}. The exactness comes from 
the fact that the functional integral (\ref{kw}) 
becomes Gaussian when $\epsilon\to0$
because the higher-loop contribution are suppresed by
the volume of the momentum integration region,
\be\label{nwhe}
e^{-\frac{1}{\hbar}S_{k-\delta k}[\vec\phi]}
=e^{-\frac{1}{\hbar}S_k[\vec\phi+\vec\psi_k]}\int{\cal D}[\vec\psi]
e^{-\frac{1}{2\hbar}\sum\partial^2S_k[\vec\phi+\vec\psi_k]
(\vec\psi-\vec\psi_k)^2}\left(1+\ord(\hbar\epsilon)\right).
\ee
Here $\vec\psi_k$ denotes the saddle point which satisfies
\be\label{saddco}
{\delta S_k[\vec\phi+\vec\psi_k]\over\delta\vec\psi}=0
\ee
and each eigenvalue of the inverse propagator 
$\delta^2S[\vec\phi+\vec\psi_k]/\delta\vec\psi\delta\vec\psi$
is positive. The appareance of the product $\hbar\epsilon$
in the higher order corrections indicates the presence of
the new small parameter $\epsilon$ which renders the one-loop
evolution equation exact. By performing the integration we find
\be\label{gel}
e^{-\frac{1}{\hbar}S_{k-\delta k}[\vec\phi]}=
e^{-\frac{1}{\hbar}S_k[\vec\phi+\vec\psi_k]}
\left(I_k[\vec\phi]\right)^{{\cal N}_d/2}
\left(1+\ord(\hbar\epsilon)\right)
\ee
where $I_k[\phi]$ is an integral over one Fourier component and
${\cal N}_d/2$ is the number of modes to eliminate in the shell 
$k-\delta k\le |p|\le k$. The factor 1/2 comes from the reality condition
$\tilde\phi(-p)=\tilde\phi^\star(p)$ and
\be
{\cal N}_d=\frac{\Omega_d k^{d-1}}{(2\pi)^d}V_d\delta k,
\ee
where $\Omega_d$ is the solid angle in $d$ dimensions.
The evolution in the cutoff is given by the 
functional finite difference equation
\be\label{sadev}
{1\over\delta k}\left(
S_k[\vec\phi+\vec\psi_k]-S_{k-\delta k}[\vec\phi]\right)=
-{\hbar\over2\delta k}Tr\ln
{\delta^2S_k[\vec\phi+\vec\psi_k]\over\delta\vec\psi\delta\vec\psi}
\left(1+\ord(\hbar\epsilon)\right)
\ee
One should bear in mind that the infrared field $\vec\phi$
is kept constant in computing this functional derivative.
Assuming that the saddle point is trivial, $\vec\psi_k=0$,
we obtain
\be\label{lexp}
\partial_kS_k[\vec\phi]=-\lim\limits_{\delta k\to0}
{\hbar\over2\delta k}Tr\ln
{\delta^2S_k[\vec\phi]\over\delta\vec\psi\delta\vec\psi}.
\ee
The limit $\delta k\to0$ is safe because the trace is
always $\ord(\delta k)$. This is not necessarily so in the
general case \eq{sadev} where the saddle point
$\vec\psi_k$ is supposed to have a finite limit as $\delta k\to0$.

The tree-level renormalization group equation, c.f. \eq{trevv}
below, can be obtained for any sufficiently smooth
action functional. But in order to convert the 
complicated functional equation
into a system of coupled differential equations for
the renormalized coupling constants
one needs an ansatz, a certain functional form for the action.
This is usually given by the gradient expansion, by assuming 
that all important terms of the action are local.
We shall consider the lowest order contributions
of this expansion which are compatible 
with rotational invariance in Euclidean $d$-dimensional space,
\be\label{kapprox}
S_k[\vec\phi]=\int d^dx\left[\hf Z_k(\vec\phi)
(\partial_\mu\vec\phi)^2+U_k(\vec\phi)\right]
\ee
where the functions $Z_k(\vec\phi)$ and $U_k(\vec\phi)$ 
are $O(N)$ invariant and depend on $\Phi=|\vec\phi|$ only. 
The running coupling constants are defined by the expansion
\be\label{tram}
Z_k(\Phi)=\sum\limits_{n=0}^\infty{z_n\over n!}\Phi^n,~~~~
U_k(\Phi)=\sum\limits_{n=0}^\infty{g_n\over n!}\Phi^n,
\ee

The qualitative features of the renormalization flow
can be seen in a simpler approximation where one
retains the local potential only,
\be                                                                  
S_k^{(loc)}[\vec\phi]=\int d^dx\left[
\hf(\partial_\mu\vec\phi)^2+U_k(\vec\phi)\right].
\ee
The corresponding Wegner-Houghton equation is
\be\label{whe}
k\partial_kU_k(\Phi)=-\hbar\frac{\Omega_d k^d}{2(2\pi)^d}
\ln\left[\left(k^2+\partial_\Phi^2U_k(\Phi)\right)
\left(k^2+\frac{1}{\Phi}\partial_\Phi
U_k(\Phi)\right)^{N-1}\right]
\ee
where $\Omega_d$ is the solid angle in dimension $d$.
Notice that the arguments of the logarithm is the inverse
curvature of the action at the minimum and each of them
is supposed to be positive, being proportional to the restoring 
force acting on the fluctuations around the vacuum. 
The solution $U_0(\Phi)$ corresponding to the initial condition 
$U(\Phi)=U_\Lambda(\Phi)$ imposed at $k=\Lambda$ 
gives the effective potential, the energy density of 
the vacum of the constrained model \eq{cmod}.

The fluctuations are stable in the $O(N)$ symmetrical phase,
\be\label{ineq}
k^2+\partial^2_\Phi U_k(\Phi)>0,~~~~\mbox{and}~~~~
k^2\Phi+\partial_\Phi U_k(\Phi)>0.
\ee
If either of these inequalities is violated then the infinitesimal
fluctuations around the given $\Phi$ become unstable. This
is the case with spontaneously broken symmetries. In fact, the arguments 
of the logarithm are decreasing monotonically as the cutoff $k$ is lowered 
but they stay positive in the symmetrical phase. The symmetry
broken phase is characterised by having a finite cutoff
value $k=k_{cr}$, where \eq{ineq} is
violated. The loop-expansion is clearly inapplicable
for $k\le k_{cr}$. Since the Goldstone modes are
the lightest excitations it is always the second inequality 
in \eq{ineq} which indicates this instability.

Let us introduce $\Phi(k)$, the curve on the plane $(\Phi,k)$ 
where \eq{ineq} is first violated as $k$ is decreased. 
The function $\Phi(k)$ is decreasing and the fluctuations 
around the constrained vacuum of \eq{cmod} with momentum 
$p<k_{cr}$ are unstable if $\Phi<\Phi_p$. This is the 
spinodal unstable phase where fluctuations
with infinitesimal amplitude are already unstable.
The Wegner-Houghton equation \eq{lexp} does
not apply any more in this regime because one has 
to take into account the non-trivial saddle points as in
\eq{sadev}. The equation \eq{lexp} should in 
principle be renounced for {\em any} value of 
$\Phi$ when $k<k_{cr}$. This is because the configurations 
whose action receives contributions from the spinodal unstable
regions are not treated in a reliable manner. 
But the strategy of the loop epxansion offers the remedy
by assuming that the amplitude of the fluctuations 
is infinitesimal. Thus one may use the simple equation 
\eq{lexp} even for $k<k_{cr}$ when $\Phi\gg\Phi(k)$ because
the fluctuations which are treated in an unrealiable manner
are strongly suppressed.

We encounter an important simplification by noting that
the loop expansion is applicable in the stable region, 
$\Phi\gg\Phi(k)$. In fact, we can use the simplest 
tree-level approximation in the stable region to obtain
an evaluation of 
$\Phi(k)$. The $k$ dependence is simply ignored in 
$\ord(\hbar^0)$, $U_k(\Phi)=U_\Lambda(\Phi)$.
The condition determining the curve $\Phi(k)$ is then
\be
k^2\Phi(k)+\partial_\Phi U_\Lambda(\Phi(k))=0.
\ee

The approach of the instability with $N=1$ has been examined numerically 
in the local potential approximation ($Z_k=1$) \cite{grg}. The spinodal 
region was considered, as well, where nontrivial saddle points appear
in the functional integration of the UV modes \cite{iir}. 
We give below an analytic derivation of the renormalized trajectory
in the spinodal instability region for a 
$N$-component field in the approximation (\ref{kapprox}).

\section{Degeneracy of the action}
The characteristic feature of the effective action
generated by the saddle points is a high level of degeneracy
\cite{iir}. To understand its implication we first note that 
the usual saddle point approximation to
the effective action or potential where all degrees
of freedom are treated in the loop-expansion is based
on an extended saddle point at zero momentum or a
gas of localized solutions which include all momenta.
The fluctuations around the saddle point are orthogonal
to the saddle point configuration and are suppressed 
by $\hbar$. But the saddle points of the infinitesimal
blocking step appear differently. Suppose that $\delta k$ is so 
small that we can neglect the dependence of the Fourier transform 
$\tilde\phi(p_\mu)$ of the field variable on the length 
$|p|$ between $k-\delta k$ and $k$. 
We may then look at \eq{nwhe} as a path integration for 
a system on the $d-1$ dimensional 
sphere, $S_{d-1}$. The $d-1$ dimensional action, 
$S^{(d-1)}_k[\phi]$ is labeled by the parameter $k$. 
It will be shown that the continuous dependence of the action 
in $k$ makes $S^{(d-1)}_{k-\delta k}[\phi]$ degenerate up 
to corrections $\ord(\delta k)$. This weak variation
of the action is enough to support the loop-expansion but
can not suppress the fluctuations which are non-orthogonal
to the saddle point for finite but small $\delta k$. This complication 
does not influence the value of the running coupling 
constants, the renormalization group flow. When the
successive elimination process of the renormalization group
method is used to carry out the path integral for some
observables in the limit $\delta k\to0$ then the following,
different strategy is employed: We simplify the path 
integration enormously by taking into account that the
effective action for each momentum shell is degenerate
in the unstable region. This degeneracy is reminescent of 
the Maxwell construction for the coexisiting phase of 
the first order phase transitions. The key element of both 
strategies, the $\ord(\delta k)$ degeneracy of the action is the
subject of this Section. 

Before embarking the general argument we remark that
the problem of finding the saddle points of the model with 
$N>1$ can be reduced to the case $N=1$ if the action is 
monotonically increasing with $(\partial_\mu\vec\phi)^2$, 
like in \eq{kapprox}. To see this let us write
\be
\vec\phi(x)+\vec\psi_k(x)=\eta_k(x){\cal R}(x)\vec e
\ee
where ${\cal R}(x)$ is a $SO(N)$ matrix and $\vec e$ the 
unit vector giving the orientation of $\vec\phi$ in the 
internal space. The saddle point is a local minimum
of the action whose dependence in ${\cal R}$ is coming by
$\partial_\mu{\cal R}$. Since 
\be
(\partial_\mu\vec\phi+\partial_\mu\vec\psi_k)^2=
\partial_\mu\eta_k\partial_\mu\eta_k+
\eta_k^2\partial_\mu{\cal R}\vec e\partial_\mu{\cal R}\vec e,
\ee
the minimum is reached by a homogeneous ${\cal R}$,
in which case $\vec\psi_k$  must be parallel to $\vec e$. 
Thus the saddle point of the model $N>1$ reduces to the 
case $N=1$, which we consider below for the sake of simplicity.

The proof of the degeneracy of the effective action
is based on the following assumptions:
\begin{enumerate}
\item the continuity of $S_k[\phi]$ 
as the functions of $k$ for a fixed configuration $\phi$ in the unstable region,
\item the infinite differentiability of 
$S_k[\phi]$ with respect to $\phi$ for any fixed value of $k$ in the 
unstable region,
\item the continuity of $S_k[\phi]$ and $\delta S_k[\phi]/\delta\phi(x)$
on the spinodal line, $\phi(x)=\Phi(k)$.
\end{enumerate}

The saddle point configuration is a solution of the non-linear
equation of motion. We have to perform the Gaussian approximation
in \eq{nwhe} for each saddle point and sum up the contributions. 
Let us denote the renormalized action resulting from the
saddle point $\psi_{k,\alpha}$ by $S_{k-\delta k,\alpha}[\phi]$, where
$\alpha$ stands for the zero modes of the saddle point.
The tree-level renormalization arises because the saddle point 
depends on the background field and \eq{gel} leads to 
\be\label{c1}
S_{k-\delta k,\alpha}[\phi]=
S_k\left[\phi+\psi_{k,\alpha}[\phi]\right]+{\cal O}(\hbar\epsilon).
\ee
In order to find the evolution of the functional derivatives
of the action one has to keep in mind that the independent variables
are the Fourier components $\tilde\phi$ and $\tilde\psi_k$. Thus one has to 
read the action as a functionnal of the Fourier components:
\be
S_k[\phi+\psi_{k,\alpha}[\phi]]=
S_k\left[\tilde\phi,\tilde\psi_{k,\alpha}[\phi]\right].
\ee
We shall suppress the index $\alpha$ in the expressions below.
The condition (\ref{c1}) then implies
\be
\frac{\delta S_{k-\delta k}}{\delta\tilde\phi(p)}[\phi]=
\frac{\delta S_k}{\delta\tilde\phi(p)}
\left[\phi+\psi_k\right]
+\int \frac{d^dq}{(2\pi)^d}\frac{\delta S_k}{\delta\tilde\psi(q)}
\left[\phi+\psi_k\right]
\frac{\delta\tilde\psi_k(q)}{\delta\tilde\phi(p)}
+{\cal O}(\hbar\epsilon).
\ee
Since $\psi_k$ is the saddle point, 
\be\label{saddle}
{\delta S_k\over\delta\tilde\psi(q)}[\phi+\psi_k]=0,
\ee
we have
\be\label{c2}
\frac{\delta S_{k-\delta k}}{\delta\tilde\phi(p)}[\phi]=
\frac{\delta S_k}{\delta\tilde\phi(p)}[\phi+\psi_k]
+{\cal O}(\hbar\epsilon)~~~~\mbox{for}~~|p|\le k-\delta k.
\eei
The higher derivatives are obtained by taking the
successive functional derivatives of (\ref{kw}) with 
respect to the Fourier components of $\phi$. The 
second derivative reads as
\bea
&~&\left[\frac{\delta^2 S_{k-\delta k}}{\delta\tilde\phi(q)
\delta\tilde\phi(p)}[\phi]
-\frac{\delta S_{k-\delta k}}{\delta\tilde\phi(q)}[\phi]
\frac{\delta S_{k-\delta k}}{\delta\tilde\phi(p)}[\phi]\right]
e^{-\frac{1}{\hbar}S_{k-\delta k}[\phi]}\nonumber\\
&=&\int{\cal D}[\psi]\left[\frac{\delta^2 S_k}{\delta\tilde\phi(q)
\delta\tilde\phi(p)}[\phi+\psi]
-\frac{\delta S_k}{\delta\tilde\phi(q)}[\phi+\psi]
\frac{\delta S_k}{\delta\tilde\phi(p)}[\phi+\psi]
\right]e^{-\frac{1}{\hbar}S_k[\phi+\psi]}\\
&=&\left[\frac{\delta^2 S_k}{\delta\tilde\phi(q)
\delta\tilde\phi(p)}[\phi+\psi_k]
-\frac{\delta S_k}{\delta\tilde\phi(q)}[\phi+\psi_k]
\frac{\delta S_k}{\delta\tilde\phi(p)}[\phi+\psi_k]\right]
e^{-\frac{1}{\hbar}S_k[\phi+\psi_k]}
\left(1+{\cal O}(\hbar\epsilon)\right).\nonumber
\eea
According to (\ref{c1}) and (\ref{c2}) we arrive at
\be\label{c3}
\frac{\delta^2 S_{k-\delta k}}{\delta\tilde\phi(q)
\delta\tilde\phi(p)}[\phi]=
\frac{\delta^2 S_k}{\delta\tilde\phi(q)
\delta\tilde\phi(p)}[\phi+\psi_k]
+{\cal O}(\hbar\epsilon).
\ee
The further derivatives can be found by induction, 
\be\label{recurre}
\frac{\delta^n S_{k-\delta k}}{\delta\tilde\phi(p_1)...
\delta\tilde\phi(p_n)}[\phi]=
\frac{\delta^n S_k}{\delta\tilde\phi(p_1)...
\delta\tilde\phi(p_n)}[\phi+\psi_k]+{\cal O}(\hbar\epsilon),
\ee
for $|p_i|\le k-\delta k$.
Let us now take the derivative of \eq{c2} with respect to $\tilde\phi(q)$:
\be
\frac{\delta^2S_{k-\delta k}}{\delta\tilde\phi(q)\delta\tilde\phi(p)}
[\phi]=\frac{\delta^2S_k}{\delta\tilde\phi(q)\delta\tilde\phi(p)}
[\phi+\psi_k]+\int\frac{d^dq'}{(2\pi)^d}
\frac{\delta^2 S_k}
{\delta\tilde\psi(q')\delta\tilde\phi(p)}[\phi+\psi_k]
\frac{\delta\tilde\psi_k(q')}{\delta\tilde\phi(q)}
+{\cal O}(\hbar\epsilon)
\ee
which, according to \eq{c3}, leads to, in the limit $\epsilon\to 0$, 
\be
\int\frac{d^dq'}{(2\pi)^d}
\frac{\delta^2 S_k}
{\delta\tilde\psi(q')\delta\tilde\phi(p)}[\phi+\psi_k]
\frac{\delta\tilde\psi_k(q')}{\delta\tilde\phi(q)}=0.
\ee
The derivative of the saddle point is supposed to be 
nonvanishing even as $\epsilon\to 0$. Since the background field $\phi$
can be chosen arbitrarily we conclude that
\be\label{c4}
\frac{\delta^2 S_k}{\delta\tilde\psi(q')\delta\tilde\phi(p)}
[\phi+\psi_k]=0~~~~\mbox{for}~~k-\delta k<|q'|\le k.
\ee
But if we take the derivative of \eq{saddle} with respect to $\tilde\phi(p)$,
we find
\be\label{c5}
\frac{\delta^2 S_k}{\delta\tilde\psi(q)\delta\tilde\phi(p)}[\phi+\psi_k]
+\int\frac{d^dq'}{(2\pi)^d}
\frac{\delta^2 S_k}{\delta\tilde\psi(q)\delta\tilde\psi(q')}[\phi+\psi_k]
\frac{\delta\tilde\psi_k(q')}{\delta\tilde\phi(p)}=0.
\ee
Because of the non vanishing of the derivative of the saddle point when 
$\epsilon\to 0$, \eq{c4} and  \eq{c5} imply that
\be
\frac{\delta^2S_k}{\delta\tilde\psi(p)\delta\tilde\psi(q)}[\phi+\psi_k]
=0~~~~\mbox{for}~~k-\delta k<|p|,|q|\le k.
\ee
One can easily find by induction, that for any $n\ge 1$ 
\be
\frac{\delta^nS_k}{\delta\tilde\psi(p_1)...\delta\tilde\psi(p_n)}[\phi+\psi_k]
=0~~~~\mbox{for}~~k-\delta k<|p_1|,\cdots,|p_n|\le k.
\ee
Therefore if $\psi$ has non vanishing Fourier components
for $|p|$ between $k$ and $k-\delta k$ and $\phi$ is the background field 
(with non vanishing Fourier components for $|p|\le k-\delta k$),
we can write,
\be\label{fondequa}
S_k[\phi+\psi]
=\sum_{n=0}^\infty\frac{1}{n!}\prod_{j=1}^n\int_{k_j}
(\tilde\psi(k_j)-\tilde\psi_k(k_j))
\frac{\delta^nS_k}{\delta\tilde\psi(k_1)\cdots\delta\tilde\psi(k_n)}[\phi+\psi_k]
=S_k[\phi+\psi_k]
\ee
What we found is that $S_k$ does not depend on the 
Fourier components $|p|=k$ in the unstable region. This result 
is independent of the choice of the saddle point $\psi_k$.
{\it The tree-level renormalization makes the action 
flat in the modes to be eliminated.}

Note that the continuity of $S_k[\phi]$ in $k$ for a fixed
$\phi$ was needed in our argument. In fact, the conclusion
about the flatness is reached for $S_k[\phi]$, the
initial condition of the infinitesimal blocking step. This
initial condition must be close to
the result of the blocking, $S_{k-\delta k}[\phi]$
according to the continuity of the renormalization group
flow in $k$. Thus our assumptions 1 and 2 yield the
flatness of the action within the instable region.

It is important to keep in mind that this flatness holds
up to corrections $\ord(\delta k)$ so the egienvalues of the
propagator in computing the functional integral \eq{nwhe}
are $\ord(\hbar\delta k^{-1})$. Taking into account that
the integration volume $\ord(\delta k)$, the one-loop
correction to the tree-level result is $\ord(\hbar)$,
as usual. Thus there is no arbitrary small parameter, 
like $\epsilon$, available in separating the tree- and 
the (one)loop-level contributions to the renormalization 
group equation. In other words, the simplification
offered by the smallness of the loop integration volume
suppresses the higher-loop contributions only and
leaves both the tree and the one-loop contributions
important.

The summation over the saddle points gives finally (with
the dependence on $\alpha$ properly displayed)
\be
e^{-{1\over\hbar}S_{k-\delta k}[\phi]}
=\sum\limits_\alpha^{N[\phi]}\int dX_\alpha{\cal F}(X_\alpha)
e^{-{1\over\hbar}S_k[\phi+\psi_{k,\alpha}]}
=e^{-{1\over\hbar}S_k[\phi+\psi_{k,\alpha_0}]}
\sum\limits_\alpha^{N[\phi]}\int dX_\alpha{\cal F}(X_\alpha),
\ee
where $X_\alpha$ stands for the zero modes of the saddle point
$\Psi_k(\alpha)$, $\alpha=1,\cdots,N[\phi]$, and
${\cal F}(X_\alpha)$ is the corresponding integration measure. 
Since the change of the number of the saddle points, $N[\phi]$ is 
necesseraly discontinuous in $\phi(x)$, and the zero mode integral
is always finite with inhomogeneous saddle points,
the assumption 3 asserts that $N[\phi]$ is constant.
The zero mode entropy can thus be ignored, giving 
\be\label{trevv}
S_{k-\delta k}[\phi]=S_k[\phi+\psi_k(\alpha_0)]+\ord(\hbar\epsilon),
\ee
where $\psi_k(\alpha_0)$ is an arbitrary saddle point.

\section{Renormalization group flow} 
We present a solution of the renormalization group equation
\eq{trevv} in this Section. The argument is based on the assumption 3.

The solution of the tree-level evolution equation 
is largly simplified by the high degree of the degeneracy,
\eq{fondequa}. This allows us to use a simple saddle
point to evaluate $S_{k-\delta k}[\phi]$. We use a plane wave
\be
\psi_k(x)=2\rho_k\cos(k\hat\omega_\mu(k) x_\mu+\theta(k))
\ee
where the unit vector $\hat\omega_\mu(k)$ and the phase
$\theta(k)$ are the collective coordinates corresponding
to the zero modes. The breakdown of the $O(d)$ rotational
symmetry gives rise to $\omega_\mu$. We tacitely 
assumed large but finite quantization box with periodic 
boundary conditions which guarantees the translational
symmetry. The translation in the direction of $\hat\omega(k)$
is no longer symmetry, the corresponding zero mode is $\alpha(k)$.

The value of the action (\ref{kapprox}) at 
$\chi(x)=\Phi+2\rho\cos(k_\mu x_\mu)$ is
\be
V_d^{-1}S_k[\chi]=U_k(\Phi)+\sum_{n=1}^\infty\frac{\rho^{2n}}{(n!)^2}
\left[nk^2\partial^{2n-2}_\Phi Z_k(\Phi)+
\partial^{2n}_\Phi U_k(\Phi)\right].
\ee
We have seen with \eq{fondequa} that the action is 
$\rho$-independent and thus the coefficient of
$\rho^{2n}$ vanishes,
\be
nk^2\partial_\phi^{2n-2}Z_k(\Phi)+\partial_\phi^{2n}U_k(\Phi)=0,
\ee
therefore we find
\bea
Z_k(\Phi)&=&z(k)\nonumber\\
U_k(\Phi)&=&-\hf z(k)k^2\Phi^2+u(k)
\eea
for an even potential.

According to the assumption 3 the
saddle point $\psi_k$ is a continous function of the 
background $\Phi$ at $\Phi=\Phi(k)$. This implies that 
$U_k(\Phi)$, $Z_k(\Phi)$ together with $\partial_\Phi U_k(\Phi)$ 
and $\partial_\Phi Z_k(\Phi)$ are continous functions, as well,
for $\Phi=\Phi(k)$ because the saddle point is found by setting 
the first derivative of the action to zero. The continuity of 
$Z_k(\Phi)$ leads to 
\bea\label{solinst}
Z_k^{unst}(\Phi)&=&Z_k^{st}(\Phi(k))\nonumber\\
U_k^{unst}(\Phi)&=&-\hf Z_k^{st}(\Phi(k))k^2\Phi^2+u(k)
\eea
where the quantities in the stable ($\Phi>\Phi(k)$) and 
the unstable region ($\Phi<\Phi(k)$) are supplied by a 
subscript $'st'$ and $'unst'$, respectively.
$Z_k^{st}(\Phi(k))$ comes from the solution of the
Wegner-Houghton equation in the stable region.

As a consequence of the continuity of 
$\partial_\phi U_k(\Phi)$ in $\Phi$, we find:
\be
\frac{\partial U_k^{st}}{\partial\phi}(\Phi(k))=
\frac{\partial U_k^{unst}}{\partial\phi}(\Phi(k))=
-Z_k^{st}(\Phi(k))k^2\Phi(k)
\ee
which results the definition of the boundary $\Phi(k)$
\be\label{instl}
Z_k^{st}(\Phi(k))k^2\Phi(k)+\frac{\partial U_k^{st}}{\partial\phi}(\Phi(k))=0.
\ee
Here both $Z_k^{st}(\Phi(k))$ and $U_k^{st}(\Phi(k))$ are given by
the solution of the Wegner-Houghton equation in the stable region.

Note the disapparence of $N$ from the equation.
One would have thought that the Goldstone modes which go
unstable earlier as the cutoff is lowered make different
boundary $\Phi(k)$ for $N>1$ and $N=1$. But it turned out that the 
instability line for $N=1$ is given by \eq{instl} instead of the
vanishing of the first factor in the right hand side
of \eq{whe}.

It is important to note that in the limit $k\to 0$ the potential in
(\ref{solinst}) recovers the Maxwell construction: the effective
potential $U_{eff}=U_{k=0}$ is a flat function of the field
between the minima $\pm \Phi_{k=0}$. The Maxwell construction 
which is a well-known feature of the vaccum of the coexisting phase
has been seen numerically in \cite{iir}. The potential \eq{solinst}
has already been reported in the approximation $Z=1$ for $N=3$ with a smooth 
cutoff by retaining the simple plane wave saddle point \cite{smnh}. But the 
smooth
cutoff makes other saddle points than the single plane wave important. Our
result, valid for any $N$ underlines the generality of the Maxwell
construction. The limit $N\to\infty$ can be used, as well, to
obtain the Maxwell construction \cite{exap}. But it remains an unjustified
step in this scheme to ignore the massive mode. The formal argument is
that the $N-1$ Goldstone modes overweight the single massive
mode. Since the saddle point appears in this latter the 
large $N$ scheme can not trace down the stabilisation 
mechanism in the spinodal phase.

Finally we mention the simple relation satisfied by the 
amplitude of the plane-wave saddle point,
\bea
\tilde\psi_k(p)&=&\rho_k(\Phi)
\left[e^{i\theta(k)}\delta(k\hat\omega(k)-p)+
e^{-i\theta(k)}\delta(k\hat\omega(k)+p)\right]\nonumber\\
\psi_k(x)&=&2\rho_k(\Phi)\cos(k\hat\omega(k) x+\theta(k))
\eea
We found that the action is flat for the field whose
magnitude is less than $\Phi(k)$. Since the action must
increase if the field goes beyond this limit we have
\be\label{modsp}
\Phi(k)=\mbox{max}_x\left\{\Phi+\psi_k(x)\right\}=\Phi+2\rho_k(\Phi)
\ee
to determine the amplitude $\rho_k$. 
First we note that the saddle point, as well as its functional derivative with
respert to the field have non vanishing limits when 
$\epsilon\to 0$, as we expected when giving the arguments leading to a flat 
action.
Then it is instructive to take
the limit $k\to0$. The saddle point becomes
homogeneous with the amplitude $2\rho_0(\Phi)$, and it can be interpreted
as the "polarization" of the vacuum due to the external field
$\Phi$. The lesson of \eq{modsp} is that the external field plus
the polarization always add up to $\Phi(0)$, the vacuum expectation
value. This complete "screening" of the external field is the result
of the degeneracy of the effective action and is in agreement with the
Maxwell construction.

\section{Correlation function}
The characteristic feature of the saddle point approximation
is the large number of zero modes, the high degree of 
degeneracy of the action. This makes the tree-level 
contribution to the correlation functions highly non-trivial.
We demonstrate this by computing the correlation function
for the constrained model \eq{cmod}
\be\label{pror}
G^\Phi_{\ell,m}(p,q)={1\over Z}
\int{\cal D}[\vec\phi]\tilde\phi_\ell(p)\tilde\phi_m(q)
e^{-\frac{1}{\hbar}S_\Lambda[\phi+\Phi]},
\ee
in the mean-field approximation where only one plane-wave 
mode is retained on a homogeneous background $\Phi$. 
The mean field approximation
consists of keeping a single variable active in the path integral
in the presence of a non-trivial background. One usually follows this
strategy on a real space lattice where a site variable is itegrated over
in the presence of a homogeneous background field. Such a mean-field
approximation is inacceptable within the spinodal unstable
phase. Instead, we suggest to
make this approximation in the momentum space. The living mode
will be a plane wave on the background of $\Phi$. The plane wave
will take large amplitude which creates an inhomogeneous ground state
supporting the phase separation. The average amplitude of the plane-wave
may serve as a measure of the "readyness" of the system
to create the phase separation with a given length scale. 

The shallow action is always a potential threat for the 
loop-expansion. This problem was circumvented in the 
computation of the tree-level renormalization group
by noting that the number of modes contributing to the
loop integrals is small, as well. This is not a sufficient
argument any more to suppress the contributions of the
fluctuations around the shallow minima for the expectation 
values, such as \eq{pror}. What happens is that the fluctuations
within the shallow part of the action do contribute 
to \eq{pror} in $\ord(\hbar^0)$. When the successive integration
of the renormalization group procedure is used for the evaluation
of the path integral in the limit $\delta k\to0$ then we may
disregard the saddle point structure and make the integration
within each momentum shell, using the appropriate effective
action. This integration is simplified in the unstable region
where the effective action is flat. Thus the system is strongly 
"disordered" in terms of the Fourier transformed variables 
$\tilde\phi(p)$ within the mixed phase due to 
the absence of the
restoring force acting on the fluctuations around the vacuum.

The $\ord(\hbar^0)$ contribution to \eq{pror} is be
obtained by restricting the integration into the regime 
where the effective action is flat and therefore
\bea
G^\Phi_{l,m}(p,q)&\simeq&\left[\int{\cal D}_\Phi[\vec\phi]
\right]^{-1}
\int{\cal D}_\Phi[\vec\phi]\tilde\phi_l(p)\tilde\phi_m(q)
\nonumber\\&=&
\delta_{l,m}\left[\int{\cal D}_\Phi[\vec\phi]
\right]^{-1}
\int{\cal D}_\Phi[\vec\phi]\tilde\phi_l(p)\tilde\phi_l(q),
\eea
where $\int{\cal D}_\Phi$ stands for the integration over the
functional space where the action is flat.
The approximation for the computation of the correlation function
consists in replacing the fields by plane waves, just as the saddle 
points, but for which we will integrate the amplitude over the whole
range where the action is flat, because of the degeneracy of the latter.
We will call $\Omega_N$ the solid 
angle in the internal space and $\alpha$ the angle between the
azimutal axis and the field $\vec\phi$. Let $k_p$ be the momentum 
of the plane wave. We will then take
\be
\tilde\phi_l(p)=r_l(p)\left[e^{i\theta_p}\delta(p-k_p)+
e^{-i\theta_p}\delta(p+k_p)\right]
\ee
and we will summ over all the possible orientations of $k_p$. 
In the internal space, we have $r_l=r\cos\alpha_l$ and the
correlation function is in this approximation
\bea
G^\Phi_{\ell,m}&=&\delta_{l,m}
\left[\int{\cal D}[k]{\cal D}[\theta]{\cal D}[\Omega_N]
{\cal D}_\Phi[r]\right]^{-1}
\int{\cal D}[k]{\cal D}[\theta]{\cal D}[\Omega_N]
{\cal D}_\Phi[r]r(p)r(q)\cos\alpha_p\cos\alpha_q\nonumber\\
&~&~~~~~~~~~~~~~
\times\left[e^{i\theta_p}\delta(p-k_p)+e^{-i\theta_p}\delta(p+k_p)\right]
\left[e^{i\theta_q}\delta(q-k_q)+e^{-i\theta_q}\delta(q+k_q)\right]
\eea
The integration over the phases $\theta$ will give a result different
from 0 only if $|p|=|q|$, otherwise $\theta_p$ and $\theta_q$ are 
independant variables and therefore remain only the terms where
we have the difference $\theta_p-\theta_q$ and one summation
over the direction of $k=k_p=k_q$, which we can take as a summation over $k$
because of the Dirac distributions. The correlation function is 
\bea
G^\Phi_{\ell,m}&=&\delta_{l,m}
\left[\int\hat{\cal D}[\vec\phi]\int d^dk \int d\Omega_N
\int_0^{\rho_p}r^{N-1}dr\right]^{-1}\\
&\times&\int\hat{\cal D}[\vec\phi]\int d^dk \int d\Omega_N\cos^2\alpha
\int_0^{\rho_p}r^{N-1}dr~r^2
\left[\delta(p-k)\delta(q+k)+\delta(p+k)\delta(q-k)\right]\nonumber
\eea
where we denoted $\int\hat{\cal D}[\vec\phi]$ the integration over
all Fourier components other then $\tilde\phi(p)$.

Let us introduce 
\be
\Gamma_d(\Phi)=\frac{\Omega_d}{d}\kappa_\Phi^d,
\ee
the volume in the Fourier space where
the saddle point amplitude $\rho_k(\Phi)$ is non-vanishing.
Here $\kappa_\Phi$ satisfies $\Phi(\kappa_\Phi)=\Phi$ and $\kappa_0=k_{cr}$. 
The correlation function is finally
\be
G^\Phi_{\ell,m}(p,q)= 2\delta_{l,m}\frac{(2\pi)^d}{\Gamma_d(\Phi)}
\frac{\rho_p^2}{N+2}\delta(p+q)
\label{ftco}
\ee
This expression differs from the correlation function given in 
ref. \cite{iir} for $N=1$ because of the integration over the 
$\ord(\hbar^0)$
fluctuations which yield a multiplicative constant only.
Note that the integration over the phase $\theta$ 
restored the translational invariance.

The integration region over the modulus is given by \eq{modsp}:
\be
\rho_p(\Phi)=\frac{1}{2}(\Phi(p)-\Phi)
\ee
and it extends over the plane waves for which the action is degenerate. 
The real-space correlation function can easily be obtained in closed form
as a function of $|\vec x-\vec y|=r$ in dimension $d=3$ for $\Phi=0$, 
\be\label{corf}
G_{\ell,\ell}^{\Phi=0}(r)={18\over g(N+2)(k_{cr}r)^3}
\left[\left({3\over r^2}-k_{cr}^2\right)\sin(k_{cr}r)
-{3k_{cr}\over r}\cos(k_{cr}r)\right].
\ee
where we took the following bare potential:
\be
U_\Lambda(\Phi)=\frac{g_2}{2}\Phi^2+\frac{g_4}{24}\Phi^4
=-\frac{k_{cr}^2}{2}\Phi^2+\frac{g}{24}\Phi^4
\ee
This function together with the ones for $\Phi=\Phi(0)/3,~2\Phi(0)/3$, and
$\Phi(0)$ is shown in Fig. 1.
for $g_2=-0.1,~g_4=0.01$. The vanishing of the
Fourier transform \eq{ftco} for $p>k_{cr}$ induces an oscillatory behaviour of 
the
diffraction integrals. The characteristic length scale of the large
amplitude oscillations due to the domain wall structure is
$\xi_{macr}=k_{cr}^{-1}$. Note the qualitative difference between the 
correlation function in the stable phase and \eq{corf}: The spinodal
instability induces a non-monotical behaviour, coming from
the trigonometrical functions, instead of the exponential decrease.

\section{Summary}
It is demonstrated that the tree-level contributions to the blocking
relation may induce a non-trivial renormalization. Whenever this 
happens the system displays instabilities and the vacuum is inhomogeneous 
and highly non-trivial. 

The assumption of the continuity of the effective action in the cutoff
within the unstable region allows us to show that the action is flat in the 
unstable region. This result is valid in each order of the loop expansion.
The flatness gives rise the Maxwell construction for the free energy.
A diffraction pattern-like correlation function was obtained in the 
mean-field approximation.

Our argument is based on the Euclidean field theory and as such can
be relevant for the description of equilibrium states. It is important to
mention that one can, in principle, repeat the steps in the Minkowski space-time.
The result is a dynamical renormalization group approach to the
formation of the condensate. A somehow similar program has already been
followed in ref. \cite{boy} which supports the flatness of the effective
potential in the spinodal unstable region.

The sharp momentum space cutoff is used in this work. It is well known that 
the strong dependence on the momentum at the cutoff induces non-local behavior 
in the real space which in turn spoils the gradient expansion \cite{grexp}. 
Though this does not happen in the tree-level, discussed
in this paper, it remains to be seen if the loop corrections
can be summed up in a consistent scheme in the exterior, stable regime.

\begin{figure}
\epsfxsize=8cm
\epsfysize=6cm
\centerline{\epsfbox{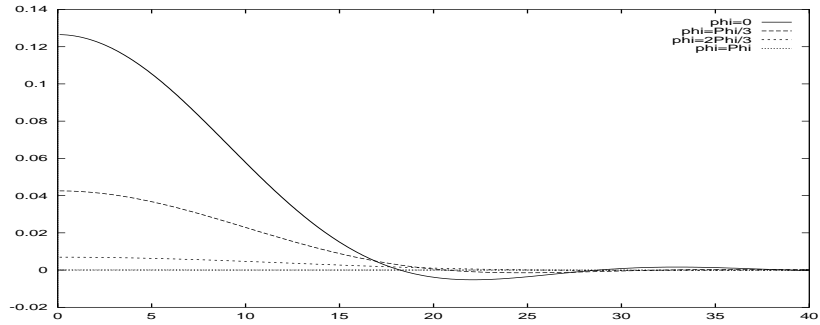}}
\caption{The correlation function in the mean-field approximation.}
\end{figure}

\end{document}